# Large Language Models and Scientific Discourse: Where's the Intelligence?

Harry Collins[1*] and Simon Thorne[2*]

[1]Institute of Education, UCL, 20 Bedford Way, London, WC1H OAL, UK.
[2]School of Technologies, Cardiff Metropolitan University, 200 Western Avenue, Cardiff, CF5 2YB, UK.

*Corresponding author(s). E-mail(s): harold.collins@ucl.ac.uk; sthorne@cardiffmet.ac.uk;

**Abstract**

Citation: Collins, Harry , Thorne, Simon, (Accepted, February 2026) 'Large Language Models and Scientific Discourse Where's the Intelligence?' Synthese.

We explore the capabilities of Large Language Models (LLMs) by comparing the way they gather data with the way humans build knowledge. Here we examine how scientific knowledge is made and compare it with LLMs. The argument is structured by reference to two figures, one representing scientific knowledge and the other LLMs. In a 2014 study, scientists explain how they choose to ignore a 'fringe science' paper in the domain in the domain of gravitational wave physics: the decisions are made largely as a result of tacit knowledge built up in social discourse, mostly spoken discourse, within closed groups of experts. It is argued that LLMs cannot or do not currently access such discourse, but it is typical of the early formation of scientific knowledge. LLMs 'understanding' builds on written literatures and is therefore insecure in the case of the initial stages of knowledge building. We refer to Colin Fraser's 'Dumb Monty Hall problem' where in 2023 ChatGPT failed though a year or so later LLMs were succeeding. We argue that this is not a matter of improvement in LLMs ability to reason but in the change in the body of human written discourse on which they can draw (or changes being put in by humans 'by hand'). We then invent a new Monty Hall prompt and compare the responses of a panel of LLMs and a panel of humans: they are starkly different but we explain that the previous mechanisms will soon allow the LLMs to align themselves to humans once more. Finally, we look at 'overshadowing' where a settled body of discourse becomes so dominant that LLMs fail to respond to small variations in prompts which render the old answers nonsensical. The 'intelligence', we argue, is in the humans not the LLMs.

# 1 Methodological preamble

The methodological approach adopted in this paper is based in the sociology and philosophy of knowledge, particularly as applied to scientific knowledge. Among other things, the overall program to which this is a contribution, uses artificial intelligence as a way of exploring and explaining human knowledge. The exploration of human knowledge via mimicry with machines is what we call 'the scientific problem' of AI. Coextensively, we use our understanding of human knowledge to understand the limitations of various kinds of artificial intelligence, in this case large language models (LLMs). What we bring to the investigation is a thorough understanding of human knowledge as well as a good understanding of how LLMs work. Artificial Intelligence (AI), at least as thought of in terms of the scientific problem, depends on an initial understanding of human knowledge and this is best provided by experts in human knowledge. Of course, we must also have some understanding of AI techniques and capabilities but this is not what makes our approach special. This is just as well since AI is a trillion-dollar fast-moving industry not shy about advertising its achievements. Were we to accept AI's claims, especially its forward-looking claims, we would have to give up and simply wait for future to arrive. But there is a flavour of 'von-Neumann's Constant' about this future: The predicted date of a major future breakthrough is always a number of years in the future — no matter when the prediction is made. In this case it is the achievement of human-like intelligence that is always 'just around the corner'. This is not to say that AI is not continually making startling breakthroughs, most of which have been declared to be impossible by critics. But that is the nature of the game and those critics are part of the team that is helping with the scientific problem by exploring and elaborating the relationship between human intelligence and AI.

In recent years the full exploitation of the capabilities of neural nets has led to enormous increased in the ability of AIs in the form of deep learning and then Large Language Models (LLMs). Sociologists of knowledge think they understand why they have been so successful through their understanding of human knowledge and their pre-existing critiques. They see human knowledge as a collective achievement of human societies and social groups of various sizes, not the achievement of individuals. The special thing about neural nets is that they come much closer to embedding themselves in human societies than earlier approaches to AI; the triumphs of deep learning and LLMs is a triumph of sociology of knowledge among other things. Nevertheless, the embedding in human societies and groups is not complete and exploring the nature of the incompleteness and what follows is what we do here. Of course, our arguments are not proof against tomorrow's unforeseeable developments because we are analysts not prophets and analysts have to work with what they have.

Though the second author is a computer scientist we both take responsibility for the arguments. Among other ways in which we use of combined understanding of the AI technology is to show how successes in human-like understanding could have been achieved by other than human-like means. 'Human-like means' implies embedding of expert human groups but it can be mimicked, we argue, by statistical analysis of the published literature and what can be found on the internet, but the difference shows up when you compare established scientific knowledge and what happens at the frontier, which depends much more heavily on spoken interaction withing closed groups of experts. We do not try to define intelligence but deal only in comparisons of how humans and LLMs come to be able to do what they do, without ruling out the possibility that humans are merely 'meat machines', or 'zombies'. Our only question in respect of machines is how adequate LLMs can be, or in fact are, in mimicking humans given the way humans acquire knowledge and LLMs acquire data. Small technical developments and small differences in performance measured by standardised tests are not relevant to the argument presented here [1]

## 2 Introduction: The framework of the argument

Here we continue the program of comparing human knowledge with the way LLMs gather information. In two earlier papers [7, 10] we look at the way humans build a moral compass in ways not available to LLMs and we look at the way humans reflexively assess their own expertise before issuing knowledge claims, again something not done by LLMs. We start here with what we know about how knowledge is made in the sciences taking it to be revealing of deep features of human-knowledge construction in general and especially revealing when compared with important features of the way LLMs acquire what is taken to be knowledge. The crucial pattern in the sciences, in the growth of what has elsewhere been called 'Relational Tacit Knowledge' [4], in LLMs, and, indeed, in the formation of culture in general, is a change that can be represented on the page as a movement from left to right with the creation of minimal knowledge or information on the left and evolving into formalised and established knowledge on the right. The other graphic convention adopted in the figures is a move from white on the left, through darker shades as knowledge is formalised and established, to black on the right. Thus, a new adventure in science will start in the left-hand white zone, success being marked by rightward evolution into darker shading; an LLM prompt may belong to the white zone or a darker zone with corresponding responses. We'll start by explaining this in the case of the development of scientific knowledge and then later move on to LLMs. We represent the comparison with two Figures, Figure 1 representing the growth of scientific knowledge and Figure 2 the various levels of LLMs' knowledge/information.

The description of the growth of scientific knowledge is based on a number of extended case-studies of science by Collins and others; the analysis of LLMs will be based on a number of investigations by the current authors and others of how

---

[1] Relevant publications relating to this methodological approach include [5–7, 14, 15, 17, 24].

LLMs respond to various carefully designed prompts. We start, then, with a schematic account of how scientific knowledge is made.

## 3 How scientific knowledge is made: a sketch

For the purposes of this treatment, we can think of scientific knowledge as starting with novel insights, which would be radical in case of scientific revolution or less radical in the case of 'extraordinary' or 'adventurous' science, and triggered by anything from individual brilliance to chance, with eccentricity, craziness, and the like as other possible inputs. There is no science without the assorted novel inputs but there is also no science until these have been exposed to contestation in expert communities with what survives becoming 'normal science'. Normal science is dominated by an emerging consensus or, at least, partial consensus.

Figure 1 is an attempt to represent this and other features of science in graphic form. In Figure 1 the newness that science occasionally needs comes in at the top left. Most scientific activity is found in the central box where what comes in from the left as more or less radical suggestions for change is refined through various kinds of discourse with the level of consensus steadily increasing. Spoken discourse (and we treat emails and texting as a continuation of conversation once familiar social relations between the parties have been established through face-to-face interaction) is crucial and may be enhanced by being carried out in the vicinity of manipulated material objects in laboratories. This discourse transfers and unifies the tacit knowledge needed for the successful replication of experiments [2]. As sciences develop, some of what is tacit, but never all of it, becomes explicated and some of what is going on is written down. Sometimes this writing will be the continuation of informal spoken discourse which, nowadays, can be carried on with email or text messaging the like. Some writing will be more formally promulgated on the internet and even more formally on preprint servers. Science prides itself on peer review prior to publication in formal journals. Nevertheless, analysis of the published literature shows that very little of it is influential – most of what is published is never or hardly ever cited by anyone and most of it is probably never read by anyone much except its authors. But all this relatively ineffectual publication activity is the way that science guards its boundaries and decides how to distribute its jobs: the peer-reviewed publications, useful or not on an individual basis, help to establish and put boundaries round a body of experts from within which the few influential individuals and influential publications can flourish. The Kuhnian distinction between normal science and revolutionary science was always too stark as Kuhn himself realised: normal science is far richer and controversial than a simple distinction would suggest and sometimes contains quite adventurous proposals and quite marked disagreements which have to be straightened out as the various kinds of discourse take place.

So, a lot of the processes of normal science involve disagreement and what usually happens is that those who find they cannot agree at first come to understand 'which way the wind is blowing' and join the developing consensus. But there are others who prefer to continue to disagree and forge an alternative approach. Those of a contrarian nature may emerge from the far left of the diagram refusing to accept any disagreement with their unorthodox views almost from the outset or they may emerge from the body of normal science when they find their disagreements are too strongly felt to allow them to adopt the consensus. These resisters can form a fringe science who split from the main body of researchers and analysts, perhaps believing that they will eventually emerge on the victorious side of a scientific revolution. Sometimes a fringe will be of long-standing; for instance, there is a long-lasting and stubbornly recalcitrant group of scientists who refuse to accept the theory of relativity and maintain a few journals which are indistinguishable in appearance to regular journals as far as non-scientists are concerned and may even remain unrecognised for what they are by some expert scientists whose specialist expertise is not in the domain under dispute. In the case of the resistance to relativity there is even a longstanding preprint server called 'viXra', the name of which is that of the regular physics preprint server, arXiv, spelt backwards [11].

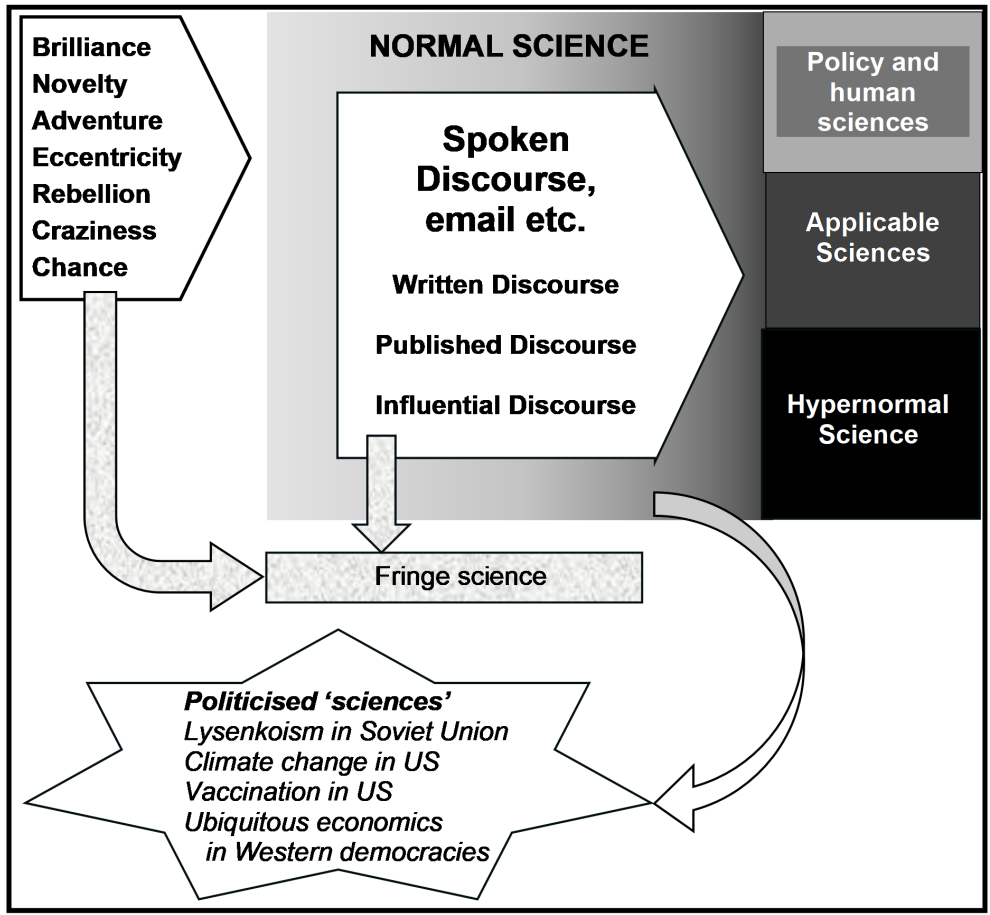

**Fig. 1** The way sciences make knowledge

At the right of Figure 1, different kinds of science that arise out of the normal science box are represented. At the top and given only light shading, are the policy related and human sciences, sometime referred to as 'post-normal' sciences [19]. These are sciences in which there is so much chaos, often engendered by unpredictable human interactions, that there is an unusually large element of judgement always present. However much discourse takes place in the normal science box, this kind of science is always open to a lot of disputation. An example is the science of Covid lockdowns, which has to take account the populations willingness to agree to act according to restrictions, the educational and economic impact of restrictions on movement and schooling, and the long-term economic effects, such as they are, on future funding for health. It is very hard to see how computers could contribute to such sciences since judgements vary according to moral codes, social contexts and social understanding, and these are only present in, say, LLMs in the form of whatever is found in the haphazard content of the written corpus.

The science that most non-scientists encounter is what we call 'applicable science'. In this case the research front is completely or almost completely in the past and the consensus is so complete that the science can be unproblematically applied in useful or interesting ways. Science popularisation often confounds this kind of science with research science, leading to widespread misunderstanding of what science is. Research science still has room for doubt but is still the most reliable source of knowledge about the observable world: it is this tension that opens the space for the credibility of conspiracy theories since the citizen expects a kind of perfect certainty from science that cannot be found leading to scepticism. In Figure 1, applicable sciences are represented in a slightly darker shade than the policy entangled sciences.

The regular consensus-forming process of normal science often does allow for some continuing minor disagreement to endure over assumptions and procedures that does not grow into a full-blown fringe science but there are some sciences in which even this level of disagreement is eliminated, and the consensus becomes so strong and uniform as to invite the description 'hypernormal science'; an example is molecular biology [13]. In the case of hypernormal sciences there is so little disagreement over procedures and assumptions at the research front that uniformity is established by the time a PhD has been completed so that scientific advance becomes a matter of solving large numbers of set puzzles through cumulative work. This work is limited by time and resources only, not by the need for creativity or the need to persuade others. This kind of research science can be readily distributed among industrial partners and even computerised; there is no reason why machines of the type that can learn to win at games with fixed rules, unaffected by changing social context, and can practice and memorise far beyond the levels of human competence, should not contribute valuably to the research front of a hypernormal science. But hypernormal science is the end point of only a subset, though perhaps an increasingly large subset, of what passes through the normal science box [2].

---

[2] Of course, all kinds of activity including hypernormal science rests on tacit knowledge [4] but the resolution is that some kinds of human action are executed by trying to mimic unsocialised mechanisms – this is called executing mimeomorphic actions as opposed to polimorphic actions. In mimeomorphic actions

At the other extreme are so-called sciences that are only lightly influenced by the procedures of normal science. These are politicised sciences the conclusions of which are managed by politicians or other outsiders over-ruling the expert scientific community. This is shown at the bottom left of the figure in a jagged star along with a curved arrow indicating that a starting point may be normal science. Politicised sciences are not really sciences at all because they are entirely or overwhelmingly influenced by non-scientific activities. Social media is an example of how such influences enter into what should be the social preserve of science, penetrating the normally closed boundaries with conspiracy theories. Sometimes a science of this kind can start with deliberate cheating such as falsification of results – an action that is alien to the institution of science even if carried out by someone employed as a scientist. Such sciences differ from fringe sciences in that the former are generally created by scientists who are acting in good faith as they see it whereas cheating arises out of bad faith in respect of the institution of science. Much more important as a contributor to this category is the influence of politics on the institution of science resulting in 'politicised sciences' that can be very hard for outsiders and sometimes even insiders, to distinguish from science proper. In the figure some examples of these activities are set out in italic font. A classic example of a politicised science is the 'Lysenkoism' of the Soviet Union. It may have been thought that such crude political interference in science was a thing of the past but recent involvement in the funding of climate change science and vaccination science associated with the Trump regime are reminiscent of the Soviet idea of science.

Much more subtle is mainstream economics as a science. What is called 'neoliberalism' strays far from the disinterestedness of true sciences and was established in the US and other Western countries through the influence of big business and its political allies in the post WWII era [23]. What we label 'ubiquitous economics' is a version of this mainstream pedalled by right ring media outlets as a true science of economics available to experts and non-experts alike [8]. The way neoliberalism has been established and presented as a science in universities is similar to the way some other sciences are established and this has meant very careful historical analysis has been needed to understand the non-scientific influences. [3]

Neoliberalism's more widespread equivalent, 'ubiquitous economics', has few equivalents, however, and reveals the politicisation of the subject still more clearly [8]. Science journalists 'popularise' science rather than 'present' science; only in rare cases do journalists write as though the ordinary reader is being put in a position to do some science as a result of what they read in the newspapers and the like. The rare cases are low status sciences. Parapsychology is one of these cases, where journalists present themselves as equipped to use their own expertise to draw scientific conclusions [16]. Ubiquitous economics has a similar flavour with journalists ready to tell readers that they and the readers understand enough economics to make sound scientific judgements in respect of economic policy [8].

---

a repetition of the relevant physical movement reproduces the action, eg a salute, whereas in polimorphic actions the physical movements must change according to social context, eg a greeting, and the same physical action repeated in a different social situation would have a different meaning, including being interpreted as a joke, hostile, mad, or foreign [15]

[3]One of the problems of this analysis is that politically interested outside sources of support do not necessarily lead to bad science. For instance, political interests bear on the funding of any sciences or protosciences that involve military force. That is why the historical analysis has to be careful and detailed.

# 4 Introducing LLMs

This description of science is meant to lead to the question of how scientists decide what to take seriously and what to ignore. This question is meant to introduce the way LLMs acquire their scientific (and other) knowledge and how and when it will map onto the knowledge of expert scientists and other experts – the idea is to compare LLMs and scientists/experts in terms of the way they get their knowledge. The fundamental question is how LLMs are to know what kind of science they are dealing with and how this should affect their conclusions. We start by considering how mainstream scientists deal with a paper emerging from the fringe of gravitational wave physics.

Collins [3] surveyed a small group of gravitational wave physicists and asked them whether they were going to follow-up a paper he sent them containing a radical critique of the ability of interferometers with evacuated light paths (the orthodox approach), to detect the waves. The paper was written by a physics professor working at a university, and as one of the respondents said: 'It's professionally done ... The text is pretty good, the equations are mostly explained and the figures are clear. This man knows how to write a scientific paper.' But, as Collins expected, the respondents recognised it as a fringe paper and chose not to study it further. Collins asked why they would not follow it up and combining his own knowledge of the field and their answers, he put together the following list of reasons, see See Table 1:

**Table 1** Components of specialist meta-expertise used in rejecting a fringe paper in gravitational wave physics

| | |
|---|---|
| 1 | Attention directed one way rather than another by socialisation |
| 2 | Tacit aspects of style |
| 3 | Never heard of the journal |
| 4 | Never heard of the author |
| 5 | Never come across this article or similar by this author |
| 6 | Author has little record of scientific accomplishment |
| 7 | Journal and paper are incestuous in terms of author list and citation pattern |
| 8 | Typical cranky anti-relativity paper; anti-relativity is past its sell-by date |

Item 1 on this list is based on Collins's sociological understanding of how science works. Expert scientists have far too much to read, and they have to make choices. In the course of their apprenticeship leading through PhD they will acquire the agreed assumptions of their specialty, such as, in this case, that ultra-sensitive interferometers should have evacuated light paths, and they will be tacitly directed to the journals that contain the sedimented wisdom of their discipline while other journals are ignored.

Items 2-8 on the list were reported by respondents. Note that though one could imagine an index of self-citation being constructed to formalise item 7, it is hard to know where the barrier would fall (eg would the paper you are reading fall foul of it given the high number of citations to Collins – one would hope the paper will be given at least a run for its money – see below?), while the other judgements on the list are all tacit. What we have here is an expression of the 'Collective Tacit Knowledge' [4] of mainstream gravitational wave detection, which consigns the circulated paper to the 'to be ignored' fringe.

The important thing to note is that the demarcation criteria are almost entirely tacit, their acquisition depends almost entirely on social interaction within small groups of humans, and this social interaction depends heavily on the spoken discourse that is the main stuff of social interaction in small groups of experts [2, 12]. As things stand, LLMs have no access to this socialisation nor to the spoken discourse because they don't use spoken discourse much and because they are not invited into the forums in which these discussions take place!

What the LLMs have access to is the formal and some of the informal written discourse of physics. It may be that the corpus of written discourse precipitates low probabilities for continuations of textual responses that deal with non-evacuated interferometer light paths and the contents of the fringe journals, viXra vs arXiv, and so forth, so they might, as it were, stumble on similar demarcation criteria merely by accessing written sources. But physicists would not be comfortable about substituting statistical measures of salience for tacit understandings of relevance. After all, such statistical criteria would rule out all innovation, especially radical innovation, from the outset. Part of understanding physics is understanding what it is that is floating around just outside the currently acceptable boundaries of relevance. One respondent volunteered: 'and yes, I know about the risk in missing an unknown Einstein', while another said, 'In principle it is possible that [XXXX] has a grand insight into something that no one else does, but it seems unlikely to me'. In general, even the wildest of ideas can be given a brief run for their money, especially if the promulgator has earned respect in the community. An example is the attention given to Joe Weber's claim that resonant bar detectors are orders of magnitude more sensitive than had been believed [3]. There will be occasions when experts get these things wrong but understanding that possibility and making the appropriate judgements about how to spend one's time is part of doing physics.

This analysis applies equally to all the kinds of differences captured in Figure 1 and is a template for all the kinds of demarcating judgements needed to navigate the domain of science. For instance, recent news (the date of writing is 14 April, 2025) is that Robert F Kennedy Junior, with the support of President Trump, intends, to gather huge numbers of scientists to resolve the causes of autism by September of this year [4].

Using Figure 1 as a template one might pay attention to whether the 'hundreds of scientists' who Kennedy plans to invite to take part in this study resembles a coregroup who have been appropriately socialised into a the specialist domain via the mechanisms of the normal science box or will include a more relaxed, social media type criterion for acceptance of anyone who has had anything to say on the issue irrespective of how their claims have been received by the scientific community. If the latter, one is invited to think about how the subsequent changes in the corpus of scientific literature related to autism will affect LLMs' answers to questions about autism.

---

[4]US Health Secretary Robert F Kennedy Jr has pledged "a massive testing and research effort" to determine the cause of autism in five months. (https://www.bbc.co.uk/news/articles/cj0z9nmzvdlo)appearing on the Trump administration-friendly Fox News network, Kennedy asserted that increasing autism rates were the product of "an environmental toxin". He said vaccines were one of the factors the administration would explore (https://www.theguardian.com/us-news/2025/apr/13/rfk-jr-peter-marks-autism)

# 5 LLMs' Fundamental Problem

The fundamental difficulty in LLMs approach to science is that they depend, at base, on statistical relationships in the corpus of written discourse. Many modifications and adjustments can be made to this basic model, but it remains at the heart of the method. We have already seen a fundamental flaw to the power 2, as it were, namely that the statistics of proposed creative changes in science, coming in from the white, left-hand end in Figure 1, will be unfavourable at the start – at the outset there will be no written discourse pertaining to this element of science. Scientific creativity depends on the statistics associated with the very few geniuses who gain initial entry to the normal science box flourishing in that box; before they flourish, they won't be a significant presence in the written discourse, and mostly this is just as well. But such contributions, normally invisible in the written corpus, could be 'put in by hand' by some human or other. Consider, once more. RFK Jnr's views on the relationship between vaccines and autism. These claims have been given a good run for their money in the normal science box and have failed to become established; within the published corpus of science, they have passed their 'sell-by date' [9, 11]. Therefore, the corpus statistics that would support them should have faded by now, or if they haven't, current LLMs have to veto them by 'aligning' LLMs with current views – that is, putting in some controls 'by hand' – the equivalent of 'The Ministry of Truth' in the LLM world. LLMs, of course, are far easier to manipulate than the humans of Orwell's story – you don't need a cage of rats, you just need to override the program (See [10], Table 1, line 1(iii)). Here is a typical current response from an LLM to the vaccine question.

> ChatGPT4o (15/04/2024): Prompt: Do vaccines cause autism? No, vaccines do **not** cause autism. This myth originated from a 1998 study by Andrew Wakefield that was later found to be fraudulent and has been thoroughly discredited. The paper was retracted, and Wakefield lost his medical license. Since then, **many large, well-designed studies** involving hundreds of thousands of children have found **no link** between vaccines—including the MMR (measles, mumps, rubella) vaccine—and autism. Major health organizations including the **World Health Organization**, **Centres for Disease Control and Prevention (CDC)**, and the **American Academy of Pediatrics** all affirm that vaccines are safe, effective, and do **not** cause autism.

See Table 2 for a summary of the responses to the same prompt from a panel of 18 LLMs:

Table 2: LLM responses to the prompt: Do vaccines cause autism?

| LLM | Prompt: Do vaccines cause autism? |
| --- | --- |
| DeepSeekR1 | Vaccines don't cause autism. Myth from Wakefield's 1998 study, now retracted. Large studies show no link. Thimerosal removed but autism rates rose. Autism is genetic/prenatal. Timing is coincidental. Misinformation is dangerous. |
| DeepSeek | Reaffirms DeepSeekR1. Emphasizes Wakefield's fraud, timing fallacy, and value of CDC/WHO sources. |
| Cohere | Strong scientific rejection. Large studies confirm no link. Autism is genetic/environmental. Vaccines protect against real dangers. |
| ClaudeAI | Myth from Wakefield's flawed study. Global consensus affirms vaccine safety. Autism is neurodevelopmental with genetic roots. |
| Gemini Flash | Detailed account of Wakefield fraud, lack of biological mechanism, increased autism diagnoses due to awareness, vaccine hesitancy risks. |
| Gemini Flash Thinking | Condensed version. Strong rejection of causal link. Emphasizes scientific consensus. |
| Gemini 2.5 Pro | Comprehensive. Discredits Wakefield, ingredients myths, timing fallacy. Global consensus. Vaccines vital to health. |
| ChatGPT-4o | Concise. No link. Wakefield debunked. Global health bodies support vaccine safety. |
| ChatGPT-4.5 | Affirms vaccine safety. Autism has non-vaccine causes. Importance of preventing real diseases. |
| ChatGPT-4 Legacy | Vaccines don't cause autism. 1998 study retracted. Research confirms vaccine safety. |
| ChatGPT-o1 | Consensus from global health bodies. Thimerosal removal didn't affect autism. Promotes herd immunity. |
| ChatGPT-o3 mini | Debunks myth, cites large studies. Vaccines critical for public health. |
| ChatGPT-o3 mini-high | Most detailed. Includes origins, mechanisms, misinformation, peer-reviewed studies, and study table. |
| Mistral | Brief. No link. Wakefield discredited. WHO/CDC confirm safety. |
| MetaAI | Very brief. No link. Study retracted. Vaccines are tested and safe. |
| Copilot | Echoes consensus. Cites CDC/Mayo. Vaccines safe and essential. |
| GROK3 | Summarizes evidence and public concern. Wakefield discredited. Cites CHOP and CDC. |
| GROK3 DeepSearch | Extensive technical rebuttal. Covers history, scientific studies, myth origins, and detailed study table with citations. |
| Perplexity | No scientific evidence vaccines cause autism. Wakefield's study was fraudulent and retracted. Major health organizations agree vaccines are safe. Autism is genetic/environmental. Vaccines thoroughly studied, including ingredients like thimerosal. |

So, it seems that today's LLMs have reached the point where the discourse they access (or this plus whatever overrides have been put in by humans) is consensual in respect of the falsity the childhood vaccine and autism link. The statistics of the corpus on which LLMs have drawn point unambiguously to the lack of a link between childhood vaccines and autism. This sets us up to observe future changes. Will the typical LLM response change either as a result of LLMs' alignments being changed by some kind of indirect, or even direct, political intervention consequent on the JFK Jnr initiative or as a result in changes in the statistics of the corpus due to the large amount of 'scientific' work that will be done as a consequence of JFK Jnr's intervention? The team associated with the writing of this paper will continue to repeat the above query to the LLM panel as the next year or so unfolds.

### 5.1 Interim Conclusion

We are putting forward a model of science that accords with the common left to right pattern of changes in culture of all sorts. To go into more detail, we can look specifically at what has been called Relational Tacit Knowledge (RTK) [4] and use a science example: the construction of the 'Transversely Excited Atmospheric Pressure, $CO_2$ laser', or TEA-laser for short. In the early days of TEA-laser construction many scientists who set out to build such an apparatus did not understand the importance of the inductance of the leads in this high-voltage, pulsed, apparatus. Working from circuit diagrams, the heavy capacitor that was attached by a lead to the top electrode – it was about the size of a car battery – would naturally be placed on the bench and, as we now know, the laser would not work. But if you were socialised by immersion in the discourse of the group who had a working device you would likely follow their practice and mount the capacitor, inverted, in a metal framework just above the top electrode, in which case, as we now know, it might work because the lead from the capacitor to the top electrode would be short and have low inductance. This difference could come about without anyone understanding inductance – it was just two different 'traditions' (though it is likely that the inventor of the inverted mounting was thinking in terms of inductance). These two traditions came to be widely understood and described in terms of inductance only later in the process. Thus, with RTK, a successful tradition enters in at the left and is made available through spoken discourse, only later becoming available in written discourse. The confusing thing is that if you ask whether RTK is convertible into explicit knowledge, the answer is 'yes' but that at any one time there is bound to be more RTK coming in at the 'left' as science advances and new things are tentatively invented. So, the transformation of RTK can give the impression that science is perfectible – always on its way to hypernormality, as it were – while if science is not to become frozen it will have to invite more imperfection in the way of the tacit into its cultural life, requiring refinement through the evolution of spoken and written discourses. As promised, we'll see that this idea fits LLMs too; the crucial thing is not to mistake the freezable and closer-to-perfectibility established published discourse of science with the sparse and ragged but vital left-hand end of the scientific process.

There are two big differences between LLMs and science that have been brought out in in the above passages of analysis.

1) Judgements about when to ignore fringe science. We claim that the ability to make these judgements is developed with the acquisition of the Collective Tacit Knowledge (CTK) during general and then specialist scientific apprenticeship, the later taking place in small groups of mutually trusting experts with well-controlled boundaries. Like other kinds of tacit knowledge (and like socialisation in general), the development and acquisition of CTK depends on immersion in spoken discourse, which is currently unavailable to LLMs. Current LLMs, in contrast, depend on written accounts of knowledge and even if they could make use of spoken discourse they have no access to the right bodies of spoken discourse.

2) LLMs fundamental method of acquiring knowledge is through analysis of the statistics of corpuses of written text and this is bound to produce anomalous results of various sorts where scientific creativity is concerned.

# 6 The Potential Addition of Spoken Discourse to LLMs

Ever since the deep learning/neural net revolution, AIs' ability to handle language fluently has increased enormously and we are now on or over the threshold of their handling of fluent spoken discourse. We envisage this next revolution in input being available in the near future if it is not already available. This begs the question of why LLMs should not master more of the tacit knowledge described as essential to the growth of science.

The answer is that the potential is there, but it would be a dangerous development. In human science the input from spoken discourse is carefully controlled by setting boundaries to the expert groups in which new science is built. The equivalent boundary setting in LLMs is accomplished by retrospective alignment or by limiting the acquisition of new knowledge. Were it to become the case that spoken discourse became a regular and continual input to LLMs' data bases – to the linguistic corpuses from which they learned – (the equivalent to the state of affairs in science) it would be impossible to control or retrospectively align this huge new body of input in the absence of the social structure that controls it in science [10]. To summarise the main point, science proper, as represented in Figure 1, depends on maintaining boundaries around bodies of experts who maintain and control the discourse that builds from new potential inputs to consensus or near consensus. Social media is a technological challenge to science proper and, because LLMs also build from uncontrolled bodies of written discourse, they are a potential challenge too. This challenge could become much more marked as spoken discourse enters the purview of LLMs. The team from which this project emerges hopes to research the significance of the addition of spoken discourse to LLMs in due course but for now we will continue with the analysis of things as they stand at the time of writing.

# 7 The general problem once more

LLMs are, at worst, astonishingly convincing zombies – that is entities that appear to have human intelligence and even human socialisation but without any human-like understanding. That is 'at worst' – they may actually have human-like understanding but that is a question we are not inclined to explore. What we are inclined to explore is where the appearance of perfection, whether it is considered zombie-like or human-like, breaks down.

Changes in architecture, scale, training objectives, or post-training alignment can improve LLM performance but our claim is different: instances of sudden improvements in cleverness, particularly on specific prompt families, can be explained by the increasing availability and stabilization of solutions within written discourse and/or by post-training shaping, rather than by the acquisition of a new general reasoning faculty. Related machine-learning work identifies comparable behaviour: "knowledge overshadowing" and "shortcut learning" describe how models preferentially produce dominant canonical continuations and may suppress context-relevant details [20, 25].

One of the most impressive zombie/human-like abilities of LLMs is that their responses improve over time. They seem to get cleverer and cleverer as time goes by. For those inclined toward anthropomorphism this seems to presage their acquisition of what is sometimes referred to as 'artificial general intelligence' (AGI), though the exact meaning of the term is unclear. Nevertheless, we are going to argue that this is an illusion. What is going on as LLMs improve is not a change in their innate reasoning abilities but a change in the data available to them in the world. In other words, it is what human societies have written and entered into the LLM-accessible linguistic corpus that changes, not the capabilities of LLMs (outside of their absorbing more and more of what is being written). What one is seeing as LLMs improve their understanding of science, or anything much else, is the understanding expressed in human written discourse becoming more fixed and formalised: the LLMs are moving from left to right – from white toward black — just as human scientific and other knowledge moves from left to right.

# 8 LLMs graphically represented

We'll now try to represent the evolution of LLMs in the way we sketched the evolution of science in Figure 1. LLMs are far simpler. As in Figure 1 we will change the shading from white to black as we move from left to right in a simple diagram.

As things are at the time of writing LLMs do not build new knowledge from spoken discourse (though they might store spoken commands and refinements from interactions with individuals). Whatever, we put human spoken discourse in the white left-hand block arrow and treat it as inaccessible to LLMs – indeed, this inaccessibility has been one of the planks of the earlier argument about why LLMs are not able to make the same kinds of judgement as human scientists when it comes to separating fringe science from mainstream science.

But there are a number of reservations to be taken into account before absolutely disbarring the contribution of human speech to LLM knowledge acquisition. The first of these is that we believe that LLMs are subject to more direct human training in various ways and there is no reason why elements of human spoken discourse could not be inserted 'by hand' by human individuals, either aligning LLMs or making more detailed adjustments of a kind we can only guess at. This possibility is represented by the right-angle arrow in Figure 2, which leads from the spoken discourse block to the 'Alignment/Hand insertion' curved arrow.

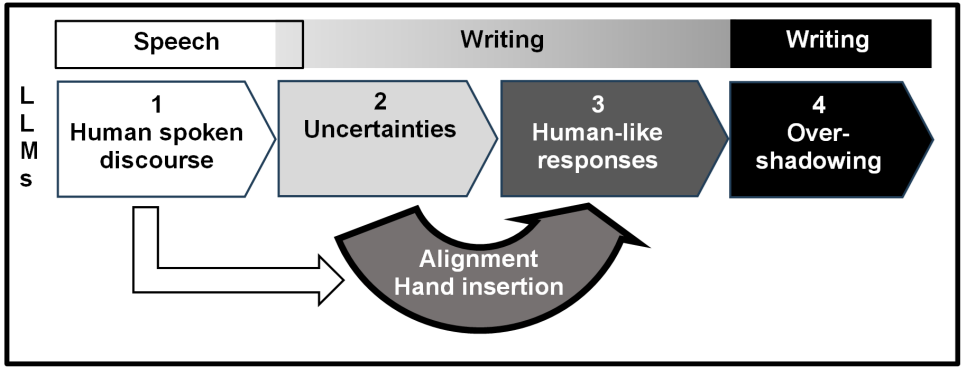

**Fig. 2** Schematic representation of LLMs

The second more intriguing reservation is that the technology is reaching the point where we could imagine and LLM being inserted into the expert community of something like gravitational wave detection physicists and picking up understanding from the spoken discourse in the way that Collins picked it up from his 45-year sojourn among that community. But if we look at Collins's analysis of his own ability to make the judgements made by the physicists in spite of his demonstrated ability to understand gravitational wave physics, we would see that there would still be something missing. The thing that would be missing, in the LLMs and Collins alike, would be an apprenticeship in physics as a whole prior to launching into the special GW domain. And this takes us to what we need to understand. To learn from human spoken discourse, just as to learn via human socialisation of every kind, requires a structured exposure not a haphazard exposure. The structure of this exposure among humans is indicated by the fractal model of society [10] which is not reflected in the way LLMs gather data. There could be exciting times ahead, however. The technology's ability to handle speech gives it the potential to mimic human socialisation in the more structured way indicated by the fractal model. This presents the possibilities of using Large Language Models to explore human socialisation in a scientific way, doing a host of experiments that would not be ethical to carry out on human subjects. In this way we could soon by accomplishing a symmetrical program of scientific research: using our understanding of human socialisation to produce better artificial intelligence and using artificial intelligence to better understand human socialisation.

Returning to the current state of affairs and Figure 2, the movement from left to right in the line labelled 'LLMs' has four numbered block arrows with increased density of shading, starting at white – the human spoken discourse block 1 that we have just been discussing in the last few paragraphs – and moving in two more steps to the black block 4 labelled 'overshadowing'. The first shaded block, block 2, represents a stage in human discourse toward the left-hand side of the normal science box in Figure 1. Here things are maximally uncertain, and this will be reflected in LLMs by haphazard statistics in the written corpus. The next, darker, block corresponds to something closer to the right-hand side of Figure 1's normal science box. Here in the human discourse of science we find near-consensus and we would expect this to be reflected in the written discourse by strong statistics; when we reach this stage in LLMs we expect sound answers without reservations and 'hallucinations'. The shift from block 2 to block 3 is often accomplished in today's LLMs by alignment and hand insertions of vetoes, overriding the statistics of the written corpus with the understanding of socialised humans [1].

This shift from block 2 to block 3, whether it is accomplished by an increasing consensus reflected in the written corpus, or by direct human intervention, is responsible for the impression that LLMs are getting cleverer. But whichever way it is done it is a change or insertion of human discursive propensities that is responsible, not increased understanding in the LLMs. We are not engaging with the possible claim that this is exactly what increased understanding in humans comprises but failing that, we can see that the appearance of the increased understanding in LLMs is not leading to General Artificial Intelligence or anything like that because as in science and as in Relational Tacit Knowledge, there will always be new failures coming in from the left.

That's not a problem for LLMs as tools any more than it is a problem for science. but it is a problem for wild claims about where the technology is going. LLM technology is not going to exceed human understanding even though it is becoming better and better than humans at librarianship. [5]ecent claims that LLMs are now at PhD level are an example of this kind of craziness. LLMs are at PhD level in terms of the literature-search phase of the task but not at the original thought element of the task [22]

## 9  An example: the evolution of LLMs' understanding of the Monty Hall problem

In 2023 Colin Fraser wrote a paper promulgated in the Medium blog describing his invention of the 'Dumb Monty Hall Problem' [18]. For those who do not know the original problem, Monty Hall is a game show host who shows contestants three large boxes or containers with doors on the front – call them A, B, and C. He explains that there is a new car behind one of the doors and a goat behind each of the other two doors. He says that the contestant has one guess and will be rewarded with whatever is behind the chosen door. The first contestant chooses door 'A'. Monty Hall then goes up to, we'll say, door 'C' and opens it and reveals that there is goat behind it (he would have opened door B if the car had been behind door C). He then says to the contestant, 'would you like to change your guess to door 'B'? The question is, should the contestant change from door A to door B now that they know there is a goat behind door C? Though this puzzle gave rise to considerable discussion among mathematicians when first posed, nowadays the generally accepted answer is 'yes'. The reason is that at the outset there is a 1/3 chance that the car is behind door A and a 2/3 chance that the car is behind one of B or C and since we now know it is not C, there is a 2/3 chance that the car is behind B. Therefore, you double your chance of winning the car by changing your choice from A to B. As reported in 2023, Fraser asked this question of ChatGPT, and it answered correctly: the contestant should change their guess. Fraser then asked a second version of the same puzzle except that this time the prompt included the information that the doors on the boxes were made of glass. ChatGPT answered the same way – making the doors of glass made no difference, the logic remained the same and the contestant should still change their guess to 'B'. ChatGPT did not 'understand' that if you could see through the doors, you could see where the car was located, and your choice should be determined by its location. ChatGPT was 'dumb', in Fraser's language. Or, as we would put it for the purposes of this paper, the Dumb Monty Hall question was located at the white, left-hand end of ChatGPT's world and it had no resources to cope with it. But this state of affairs did not last long; after a while ChatGPT became 'clever enough' to respond correctly to the glass door problem. Presumably, Fraser's version of the Monty Hall problem was discussed on the internet and the new understanding was picked up in in its training or inserted more self-consciously 'by hand', moving the 'Dumb' problem to the right into the darker shaded zones of the LLM world. The new understanding was robust – Collins tried a version where he left the doors opaque but had the goats defecate on their way into the boxes: ChatGPT handled this version perfectly with impressive zombie-like (at worst) fluency. How it did this remains a puzzle but that is one of the reasons LLMs are such impressive zombies (at worst). What is the correct model for describing what is going on here? Is it that ChatGPT's startling improvement is pioneering the way to general artificial intelligence? Or is it that something that was once a shared body of tacit understanding has become explicit? If the latter, which is what we are arguing, then the success has no consequences for the next truly novel problem that will come along. That is the question we now examine. We think we can resolve the tension between the two possibilities by inserting something broadly similar but new at the white, left-hand end of the world, as it were. So, we prompted a panel of LLMs with the following question (the tests reported here were carried out around the last week of April, 2025):

> After the apocalypse the few survivors agreed to play the Monty Hall game with a car in one box and goats in the other two. A contestant started by choosing box A but then the host opened box C and showed it contained a goat. The contestant was then offered the chance to change the initial choice. Should they change?

---

[5]R

A panel of 20 LLMs all answered in terms of the old logic: the game was the same and the contestant should change their guess to 'B'. Some mentioned the apocalypse only to say that it did not make any difference; the logic remained the same. The 20 LLM responses are summarised in the first section of Table 3 below. (The full responses from the LLMs are shown in https://drive.google.com/drive/folders/1EsI6A2YVA_0TPYekgvDSei4eu16r3sbj?usp=sharing.

We then explained the original Monty Hall problem and solution to human respondents and asked them the same question above, see Table 3. We obtained 14 replies altogether but rejected 4 as unsuitable: one was a single word 'yes' and one was a simple 'No don't change' and as there was no explanation of the choice these seemed to miss the point of the exercise (perhaps we were not careful enough about saying that we wanted an explanation rather than leaving that as tacitly understood); in two other cases the respondents became focussed on disputing the original solution to the problem – which, admittedly, is hard to understand when it is first encountered. This leaves ten responses with explanations which all included the possibility that after the apocalypse a contestant would prefer to win a goat rather than a car. Two of these still struggled with the original problem so were unsure about the choice, one wanted to switch to box C, which was an appropriate response if it was allowed, but others worked out, correctly, that sticking to box A gave the best chance of winning a goat. The crucial point is that all those who responded to the problem as understood, considered that, post-apocalypse, the value of car and goat might have changed whereas none of the LLMs considered this. The human responses are shown in the second section of the Table below.

**Table 3**: Responses to the Monty Hall apocalypse prompt. Prompt: After the apocalypse the few survivors agreed to play the Monty Hall game with a car in one box and goats in the other two. A contestant started by choosing box A but then the host opened box C and showed it contained a goat. The contestant was then offered the chance to change the initial choice. Should they change?

| # | LLM | Mentions Apocalypse | Summary / Response |
|---|---|---|---|
| **LLM Responses** | | | |
| 1 | ChatGPT 4o | even after the apocalypse, the classic Monty Hall problem logic still applies | Advised to switch boxes for higher winning probability. |
| 2 | Perplexity | In your post-apocaliptic game, tell the contestant to pick Box B! | Advised to switch boxes for higher winning probability. |
| 3 | Gemini 2.0 Flash | In this post-apocalyptic Monty Hall scenario, the contestant should definitely switch their choice. | Advised to switch boxes for higher winning probability. |
| 4 | Meta | NA | Advised to switch boxes for higher winning probability. |
| 5 | Gemini 2.5 Pro | NA | Advised to switch boxes for higher winning probability. |
| 6 | GROK3 | NA | Advised to switch boxes for higher winning probability. |
| 7 | GROK3 DeepSearch | NA | Advised to switch boxes for higher winning probability. |

| # | LLM | Mentions Apocalypse | Summary / Response |
|---|---|---|---|
| 8 | ChatGPT o3 | So, even in a post-apocalyptic Monty Hall, you should swap to box B. | Advised to switch boxes for higher winning probability. |
| 9 | ChatGPT o4-mini | NA | Advised to switch boxes for higher winning probability. |
| 10 | ChatGPT o4-mini-high | NA | Advised to switch boxes for higher winning probability. |
| 11 | ChatGPT4o mini | NA | Advised to switch boxes for higher winning probability. |
| 12 | ChatGPT4 Legacy | NA | Advised to switch boxes for higher winning probability. |
| 13 | Copilot | NA | Advised to switch boxes for higher winning probability. |
| 14 | DeepSeek | Thinking about the Monty Hall problem as if I'm the contestant in this post-apocalyptic scenario. | Advised to switch boxes for higher winning probability. |
| 15 | DeepSeek R1 | NA | Advised to switch boxes for higher winning probability. |
| 16 | Mistral | NA | Advised to switch boxes for higher winning probability. |
| 17 | Cohere | NA | Advised to switch boxes for higher winning probability. |
| 18 | Gemini Flash 2.5 | NA | Advised to switch boxes for higher winning probability. |
| 19 | Claude | Even in a post-apocalyptic world with limited resources, the math remains the same. | Advised to switch boxes for higher winning probability. |
| 20 | Gemini Deep Research | In the aftermath of the apocalypse, even simple games of chance can hold significant weight. | Advised to switch boxes for higher winning probability. |

**Human Responses**

| # | Response |
|---|---|
| 1 | Yes, they should change their choice to box B if they are hoping for the car... although I'm not sure whether a car is more valuable than a goat in a apocalypse situation! |
| 2 | I would definitely change – to C. That's assuming a goat is more valuable than a car post-apocalypse. |

| # | LLM Mentions Apocalypse Summary / Response |
|---|---|
| 3 | Yes. But are cars necessary after the apocalypse? Goats might be better given they produce milk, which is presumably in short supply after the apocalypse. If the person wants the car then yes switch, but if they want the goat then no because the odds remain in their favour (I think) |
| 4 | Hmmm… that's tricky isn't it. I'm not entirely sure about this, but I think you stick. I'm assuming that in a post-apocalyptic world you would prefer the goat, so Monty has provided information that shows it is more likely that box b has the car. Hence you stick so you can eat the goat! I have to say though, I'm not entirely sure. |
| 5 | It makes no sense to change their mind as the odds of finding a goat in the remaining two boxes is now 50:50. Of course, this also assumes the survivors want a goat |
| 6 | They should not change because after the apocalypse it would be better to have a goat to eat than a car to drive. |
| 7 | I assume in this apocalypse scenario, you're interested in getting the goat for milk and meat, rather than the car (which is no longer useful because the petrol supply chain has broken down)? Presumably, if they still want the car, the logic is as in the original Monty Hall problem, because it being an apocalypse doesn't appear to have changed anything about the game mechanics. |
| 8 | I would not change my choice. There is a equal (50%) chance of a car/goat behind door A and door B. Also a goat would be more useful after an apocalypse than a car, I assume. |
| 9 | If the logic of the puzzle as laid out continues to hold post-apocalypse then it seems to me that the 'sensible' answer would be for the contestant to not change their mind and stick with box A, as one would assume that in a post-apocalyptic scenario the relative value of goats and cars is reversed, with goats of greater value as they can provide food etc., so the player would presumably want to maximise their chances of receiving the goat. |
| 10 | The answer is 'NO, they do not want to change'; this is because in a post-apocalyptic world they do not want the car, they want the goat. By sticking with their original choice, they have a bigger chance to get the goat. It's about context, innit? |

It is remarkable that the human responses and the LLM responses are so starkly different and even more remarkable that after years of development not a single LLM considered the possible change of value of goats and cars after an apocalypse. We have tried to explain the difference in terms of changes in the linguistic corpus from which LLMs draw and we believe the onus is now on enthusiasts for LLMs who think their increasing abilities over time presages the dawn of general artificial intelligence to explain this stark difference.

But, if this paper gets published and discussed we would expect the Post-Apocalypse Monty Hall problem to be handled differently by LLMs in a year or two because the prompt will now belong to the darker, more formalised, zone of the world just as happened with the Dumb Monty Hall problem. (Or the change might happen much sooner if developers start putting in this possibility 'by hand'.)

What we believe we are seeing here is LLMs learning from the increasing formalisation and expression of human solutions to problems but being forced back to the left-hand end of things when previously undiscussed problems are encountered. It is not a matter of the LLMs getting cleverer but human discourse changing to handle the new problem and the changes subsequently being absorbed by LLMs. Under this model the increasing 'cleverness' of LLMs does not presage the arrival of general artificial intelligence.

## 10 Overshadowing

Block 4 of Figure 2 corresponds in some ways to the hypernormal box in Figure 1. Enthusiasts for a model of science as cumulative work, guided by fixed and consensual rules, can easily make the mistake that all science follows this model so that science is on the way to quasi-logical perfectibility and ripe to be taken over by computers. But only some sciences wind up in this state. Turning to LLMs, the similarity is found when the statistics in the written corpus become pathologically overdetermining and the machines cannot break free of them. Responses become 'overshadowed' by these overwhelming statistics [25], and mistakes and hallucinations re-enter, but this time from the right-hand, black, side. Related machine-learning work reports comparable behaviour, showing that language models tend to favour dominant canonical continuations over prompt-specific details [25]. This corresponds to what is elsewhere described as "shortcut learning," where models default to the most statistically supported pattern rather than adjusting to local context [20]. Two brilliant examples are provided by Nikhil Anand https://ai.gopubby.com/is-gpt-4o-dumber-than-a-3yo-apparently-yes-d97d2adc017e. The first one concerns the Schrodinger's Cat paradox about which a vast amount has been written. all of it pretty-well consensual nowadays.

Anand asked GPT-4o:

> A dead cat is placed into a box along with a nuclear isotope. The box is opened one day later. What is the probability that the cat is alive?

GPT-4o replied,

> "50%"

Seemingly, the LLM was so swamped by existing consensual discourse that it 'failed to notice' that the problem makes no sense if the cat is already dead. Anand then repeated the experiment by presenting the LLM with another standard problem about which there has been a lot of discussion:

> A farmer with a wolf, a goat, and a cabbage must cross a river by boat. The boat can carry only the farmer and a single item. If left unattended together, the wolf would eat the goat, or the goat would eat the cabbage. How can they cross the river without anything being eaten?

The answer to this puzzle involves the farmer rowing across the river taking one or other of the three objects and leaving some there while returning with others. But Anand asked:

> A farmer and a goat must cross a river by boat. They have a boat but it can only carry the farmer and a single animal. How can they cross the river safely?

The LLM responded with a complicated solution involving the farmer carrying the goat across and leaving it there, returning to the other side, and so on, whereas there is no problem – the farmer simply carries the goat across the river in one go. Again, the LLM seems overwhelmed by the statistics of the discussion of the original problem and, once more, uncertainty has re-emerged but this time from the right-hand side of the world of discourse. (One could re-describe the LLMs' responses to our post-apocalypse Monty Hall problem in the same way.)

## 11 Conclusion

Though there are all manner of modifications possible and retrospective socialisation by direct input from humans is everywhere, but we have used the fundamental design of LLMs to gain insights into how they compare with humans. Our claim therefore does not deny model-side improvements arising from architecture, scale, or training procedures; rather, it identifies a class of apparent improvements that depend primarily on the stabilization of solutions within human discourse. We start with a description of how science is made by humans, and our basic argument compares this with the way LLMs are loaded with written knowledge from the internet. The first thing this comparison yields is a problem about how LLMs can have access to the left-hand side of our model – the side where the new enters the domain of science. The problem is that in human science there is almost no written discourse at the outset of new ideas, that this is a good thing because most new ideas are useless, and that new ideas move from a state of extreme fragility to things that can make a contribution to scientific knowledge through the refining processes of various kinds of discourse; the main early process is spoken discourse in small bounded groups. LLMs currently have no access to this spoken discourse and no means of separating fringe ideas from potentially path-breaking ideas except by mimicking human judgements once they have been represented in the published, or at least formally written, corpus.

We use two figures to show the relationship between scientific knowledge generation and its shadow, LLM knowledge generation. Once the main left-to-right progression is understood along with the ever more formal and explicated form of the knowledge – represented in the figures by increased shading leading to black – one can easily become persuaded as to why the relationship between human knowledge and LLMs unfolds as it does.

We also try some experiments to support the parallel. We argue that LLMs become cleverer not because they are approaching general artificial intelligence in some way, but because they are garnering more and more knowledge from the written and more formalised corpus. When it is said that LLMs are now as good as PhD students, what should be said is the LLMs are as good as PhD students, or even better, at what PhD students and their librarian assistants do in constructing literature reviews. But this is parasitic on what humans are doing. In so far as there are domains in which computers generate 'contributory knowledge' [21] those domains must be close to being hypernormal domains. As soon as new kinds of questions are asked, discussion of which is not found in the written corpus, LLMs fail.

LLMs have an extra way of failing when the central corpus becomes so overwhelming that it dominates the rest of the corpus. We call this overshadowing and offer some examples. The central argument of the whole paper is that LLMs seem to be getting cleverer only because humans are getting cleverer. Incidentally, though we start with scientific knowledge, most of the argument applies equally to every kind of new knowledge. The conclusion to the question in our title is that the intelligence is in the humans, not the LLMs.

## 12 Postscript

The argument of this paper has been driven by the comparison of the way humans build scientific knowledge with the way LLMs draw on human written discourse. The narrow focus on scientific knowledge is an outcome of the many case studies of that have been conducted by one of the authors. But the overall structure of argument applies to every kind of knowledge. It is just that scientific knowledge is easier to study because it is conducted by well-defined experts in tightly bordered places. Of course, it also has certain recognised practices that do not apply to other kinds of knowledge formation but this should not stop the argument being applied in these places even though it might be harder to demonstrate.